\documentclass[epj,twocolumn]{webofc}
\usepackage{amsmath}
\usepackage[varg]{txfonts}   
\usepackage{hyperref}
\hypersetup{
    pdftitle = {Azores Proceedings},
    pdfsubject = {Azores Proceedings},
    pdfauthor = {Filipe Pereira},
    colorlinks = true,
    linkcolor = black,
    citecolor = black,
}
\usepackage{graphicx}

\wocname{epj}
\woctitle{Seismology of the Sun and the Distant Stars 2016}
\begin{document}
\title{SIGS - Seismic Inferences for Glitches in Stars}
%

\author{\firstname{L. Filipe R.} \lastname{Pereira}\inst{1,2}\and
        \firstname{João P.S.} \lastname{Faria}\inst{1,2} \and
        \firstname{Mário J.P.F.G.} \lastname{Monteiro}\inst{1,2}
}

\institute{Departamento de Física e Astronomia, Faculdade de Ciências da Universidade do Porto, Porto, Portugal
\and
            Instituto de Astrofísica e Ciências do Espaço, Universidade do Porto, Portugal
          }

\abstract{%
The increased amount of high precision seismic data for solar-like stars calls for the existence of tools that can extract information from such data. In the case of the study of acoustic glitches there are no publicly available tools and most existing ones require a deep knowledge of their implementation. 
In this work a tool is presented that aims to both simplify the interaction with the user and also be capable of working automatically to determine properties of acoustic glitches from seismic data of solar-like stars.
This tool is shown to work with both the Sun and other solar analogs but also shows that are still severe limitations to the methods used, when considering smaller datasets.
}
\maketitle
%
\section{Introduction}
\label{intro}

Acoustic glitches, such as the base of the convective zone (therein BCZ) and the helium second ionization zone (therein HeII), are regions in the interior of a star where the sound speed suffers an abrupt variation due to the sharp change in the internal structure. Since the eigenfrequencies of a star are directly related to the local sound speed, these glitches cause perturbations in them. These perturbations exhibit an oscillatory behavior when considering frequencies of sequential radial order \cite{gough_1988, vorontsov_1988, gough_1990}. It is thus possible to characterize some properties of these acoustic glitches by isolating and studying their contribution to the eigenfrequencies.

This oscillatory signature from the acoustic glitches has been extensively studied for the Sun to determine the location of the two previosly mentioned acoustic glitches, BCZ and HeII \cite{monteiro_1994, monteiro_2005} and it has also been applied to solar-like stars \cite{monteiro_2000, houdek_2007, faria_thesis, mazumdar_2014}.

However, the tools existent nowadays to perform this study are not publicly available and require an intimate knowledge of the implemented methods in order to achieve good results. The work presented here aims to improve upon existing methods in order to make available a numerical procedure that is both more precise than the current methods and at the same time highly user-friendly so that it can be openly distributed to the community. Starting from the work developed by Faria \cite{faria_thesis}, which in turn was adapted from the works of Monteiro et al. \cite{monteiro_1994}, Monteiro and Thompson \cite{monteiro_2005} and Mazumdar et al. \cite{mazumdar_2014} the method was improved upon and an automatic pipeline was implemented to achieve robust results independent of user intervention.

The presented tool is validated initially using observational data from the Sun and afterwards it is applied to both stars from the 16 Cygni binary and also to 10 solar-like stars from chosen from the sample of Mazumdar et al. \cite{mazumdar_2014}.

In the next section both methods implemented in the tool are described. Then in section \ref{sec:validation} the code is validated using data from the Sun. In section \ref{sec:automatic} the automatic implementations of the code are described. In section \ref{sec:results} the results obtained for the mentioned stars are shown and finally in section \ref{sec:conclusions} the results are discussed and some conclusions are drawn.

\section{Methods} 
\label{sec:methods}

The program presented in this work is comprised of two similar but independent methods capable of isolating the oscillatory signature from the acoustic glitches and then determining some of their properties. 

\subsection{Method A} 
\label{ssub:method_a}

Method A isolates the signature in the frequencies themselves by iteratively removing a smooth component from the frequencies and then fitting the residuals to the free parameters from equation
\small
\begin{equation}
\begin{aligned}
	\delta\nu \simeq & \> \nu_s \> +\> \\
	+ & A_{BCZ} \left( \frac{\nu_{r}}{\nu} \right)^2 \textnormal{cos} (4 \pi \tau_{BCZ}\nu + 2 \phi_{BCZ}) \> + \\
	+ & A_{HeII} \left( \frac{\nu_{r}}{\nu} \right) \textnormal{sin}^2( 2 \pi \beta_{HeII} \nu ) \textnormal{cos} ( 4 \pi \tau_{HeII} \nu + 2 \phi_{HeII} ).
	\label{eq_freq}
\end{aligned}
\end{equation}
\normalsize
adapted from Faria \cite{faria_thesis} where $A_{BCZ}$, $\tau_{BCZ}$ and $\phi_{BCZ}$ correspond to the amplitude, acoustic depth and phase of the signal from the base of the convective zone and $A_{HeII}$, $\tau_{HeII}$, $\phi_{HeII}$ and $\beta_{HeII}$ correspond again to amplitude, acoustic depth and phase and finally, the acoustic width of the signal, this time due to the helium second ionization zone.
$\nu_s$ is the smooth component removed from the frequencies which is obtained independently for the $N_l$ frequencies of each angular degree $l$ by fitting a polynomial of degree $N_l-1$ with third derivative smoothing \cite{monteiro_1994}. This smoothing can be controlled by defining a smoothing parameter $\lambda$ which can be controlled by the user but is by default determined automatically by the code.

After removing the smooth function from the frequencies, the residuals are fitted to Equation \ref{eq_freq} using the PIKAIA genetic algorithm \cite{pikaia} which is a global minimization algorithm that utilizes concepts inspired by the processes of evolution by natural selection. This method is advantageous compared to hill climbing methods because it doesn't require an initial guess and instead it explores the user-defined range of possible values for each of the free parameters and should, given enough time, converge to global minimum in the parameter space. Minimization is achieved by initiating a population of sets of possible solutions chosen randomly from the interval and then evolving then for a specified number of generations. Through the generations the best sets from the population are chosen to move forward by using a standard chi squared fitness function and after the evolution process the best set of parameters is chosen as the correct one.

\begin{figure}[h]
\centering
\includegraphics[width=\hsize,clip]{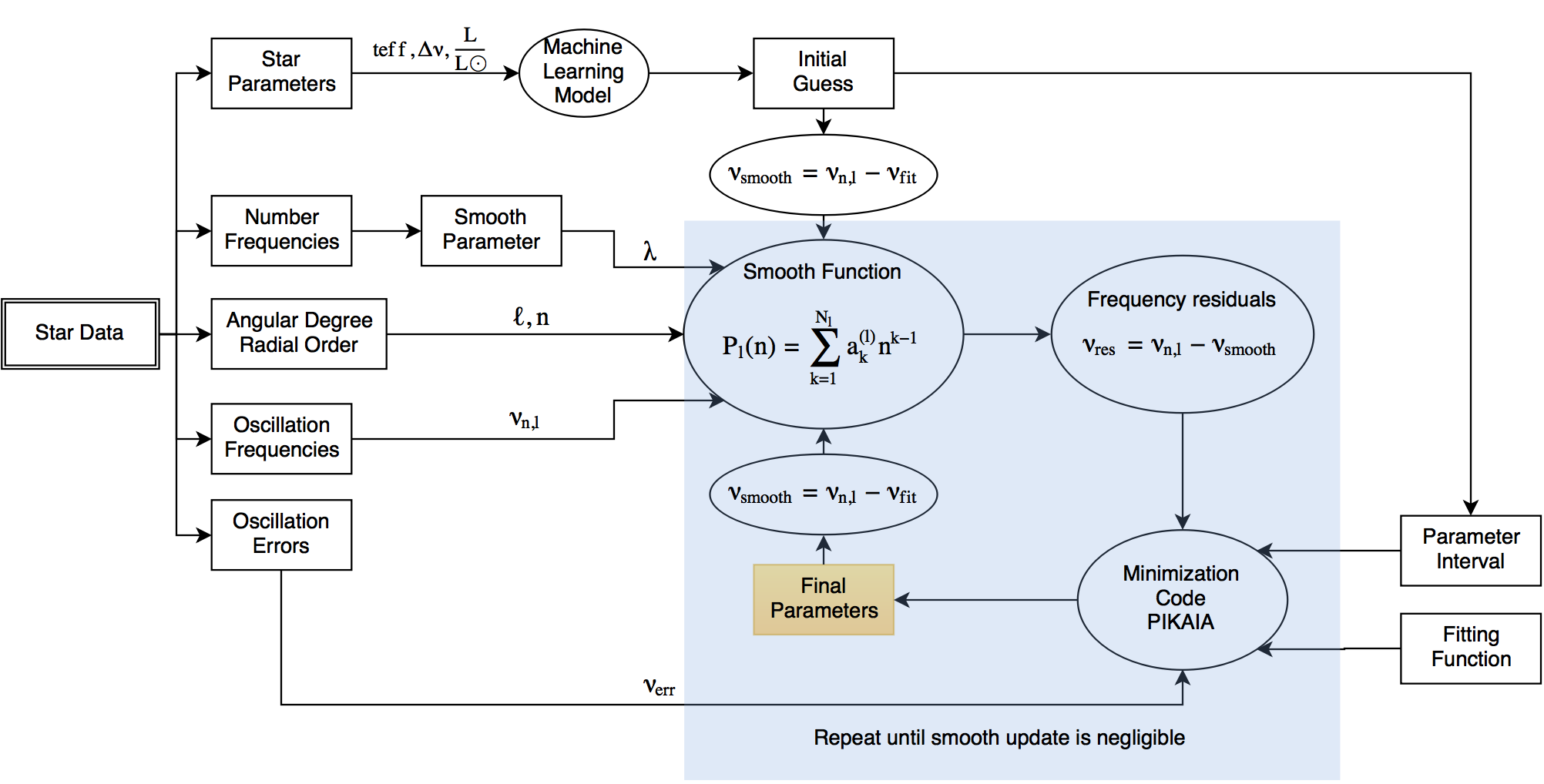}
\caption{Diagram of the workflow of method A of SIGS which determines the acoustic depths of the base of the convective zone and helium II ionization zone of a star by extracting the signal caused by these glitches directly from the oscillation frequencies of the star.}
\label{work_freq}       
\end{figure}

The process of executing the smooth removal and fitting the residuals to the equation is repeated until the smooth function is perfectly characterized and convergence is said to be achieved. Figure \ref{work_freq} shows a diagram of the general workflow of this method

\subsection{Method B} 
\label{ssub:method_b}

In method B, instead of using the frequencies, the method searches for the signature of the glitches in the second differences \cite{gough_1990}. 

In this method the smooth component is only removed once and the functional form of this component is a polynomial of a degree up to three depending on the amount of data available for the fit.

After removing the smooth function the procedure is similar to the other method with the exception that there is no iteration on the smooth component and the functional form of the signal is taken to be
\small
\begin{equation}
\begin{aligned}
	\delta \Delta_2 \nu \simeq & \> \Delta 2 \nu_s \> +\> \\
	+ & A^*_{BCZ} \left( \frac{\nu_{r}}{\nu} \right)^2 \textnormal{sin} (4 \pi \tau_{BCZ}\nu + 2 \phi_{BCZ}) \> + \\
	+ & A^*_{HeII} \left( \frac{\nu}{\nu_{r}} \right) \textnormal{exp}\left[ - \beta^*_{HeII} \left( \frac{\nu_{r}}{\nu} \right)^2 \right] \textnormal{sin} ( 4 \pi \tau_{HeII} \nu + 2 \phi_{HeII} )
	\label{eq_diff}
\end{aligned}
\end{equation}
\normalsize
which is very similar to the one from the frequencies with the same seven free parameters but a different smooth function as mentioned previously. This expression was adapted from the work of Mazumdar et al. \cite{mazumdar_2014}, which in turn was adapted from the work of Houdek and Gough \cite{houdek_2007}

\begin{figure}[t]
\centering
\includegraphics[width=\hsize,clip]{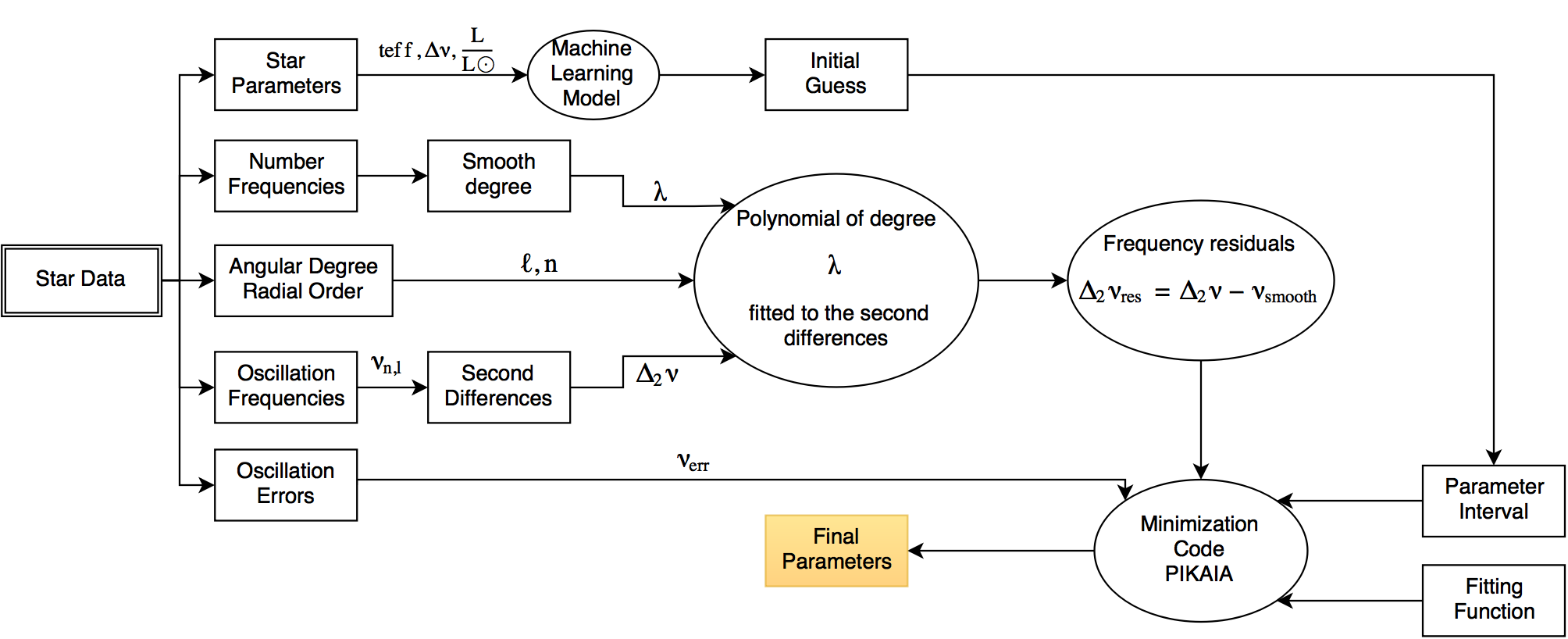}
\caption{Diagram of the workflow of method B of SIGS which determines the acoustic depths of the base of the convective zone and helium II ionization zone of a star by extracting the signal caused by these glitches from the second differences of the frequencies of the star.}
\label{work_diff}       
\end{figure}

Figure \ref{work_diff} shows the workflow of method B.

\subsection{Monte Carlo Simulations} 
\label{sub:monte_carlo_simulations}

To add robustness to both methods and estimate the errors in the measurements of the parameters we perform Monte Carlo simulations. We generate alternative data sets by taking the original frequencies and changing them by a value sampled from a normal distribution with standard deviation equal to the uncertainty of the original frequency.

The methods are then used on all the new data sets to obtain a distribution of values for the final parameters.



\section{Validation} 
\label{sec:validation}

To validate the implementations of both methods we apply them to both low-degree solar frequencies \cite{bison_frequencies} and then compare the results with that of previous works.

\begin{figure}[h]
\centering
\includegraphics[width=\hsize,clip]{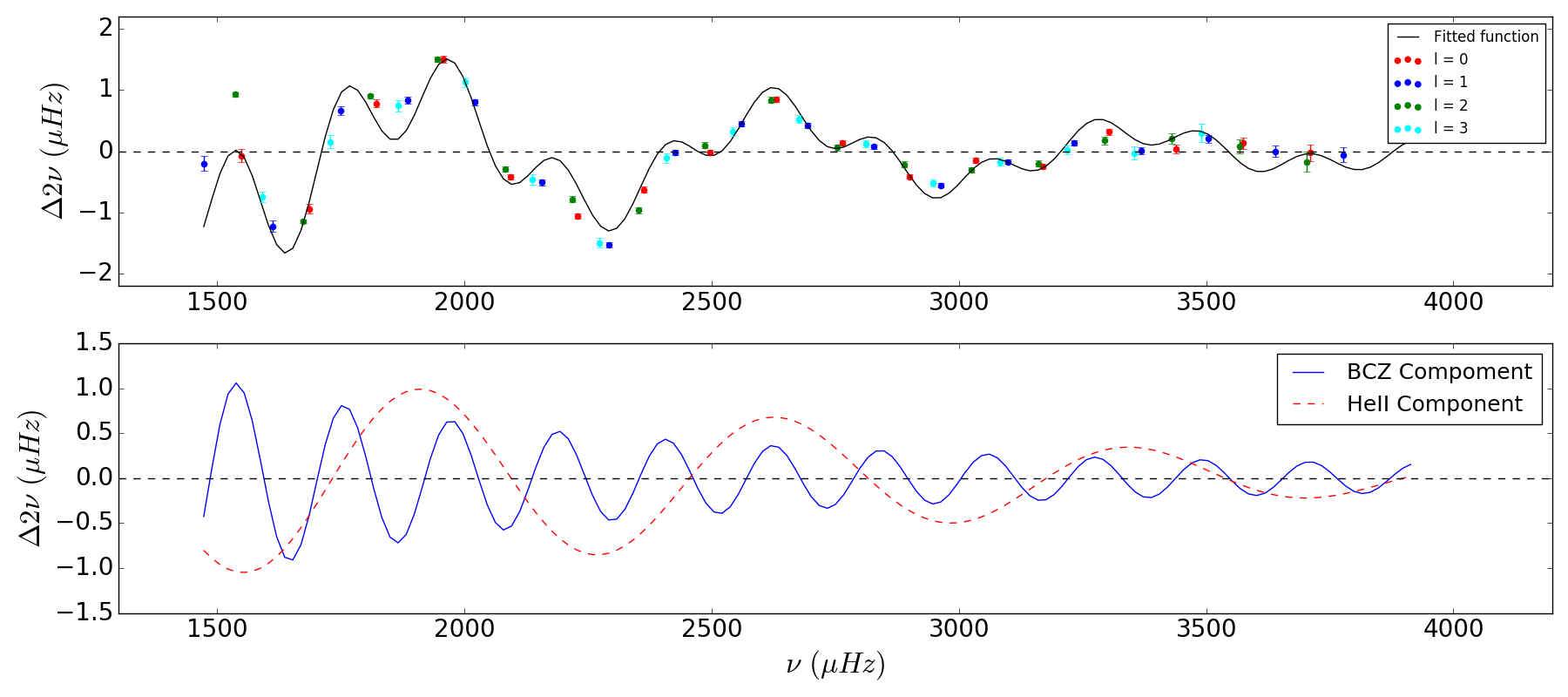}
\caption{The top panel shows the fit of the second differences from the solar frequencies to Equation \ref{eq_diff} using Method B. The color scheme detailed in the legend is used to differentiate between the degrees l of the second differences. The bottom panel has the signal components from both the BCZ and HeII resulting from the fitting of the upper panel.}
\label{fit_sun_diff}       
\end{figure}

The results obtained for both methods are summarized in Table \ref{sun_res}.
\begin{table}[h]
\centering
\caption{Comparison of the location of the Sun's acoustic glitches determined using observational data.}
\begin{tabular}{c | c c | c c}
& \multicolumn{2}{c}{This Work} & \multicolumn{2}{c}{Previous Works} \\
& $\delta \nu$ & $\Delta 2 \nu$ & $\delta \nu$ & $\Delta 2 \nu$ \\ [0.5ex]
\hline
$\tau_{BCZ} (s)$  & 2281.59 & 2313.71 & 2337 & 2273 \\
$\tau_{HeII} (s)$ & 685.25 & 692.29 & 649 & 707 \\
\end{tabular}
\label{sun_res}
\end{table}
\begin{figure}[h]
\centering
\includegraphics[width=\hsize,clip]{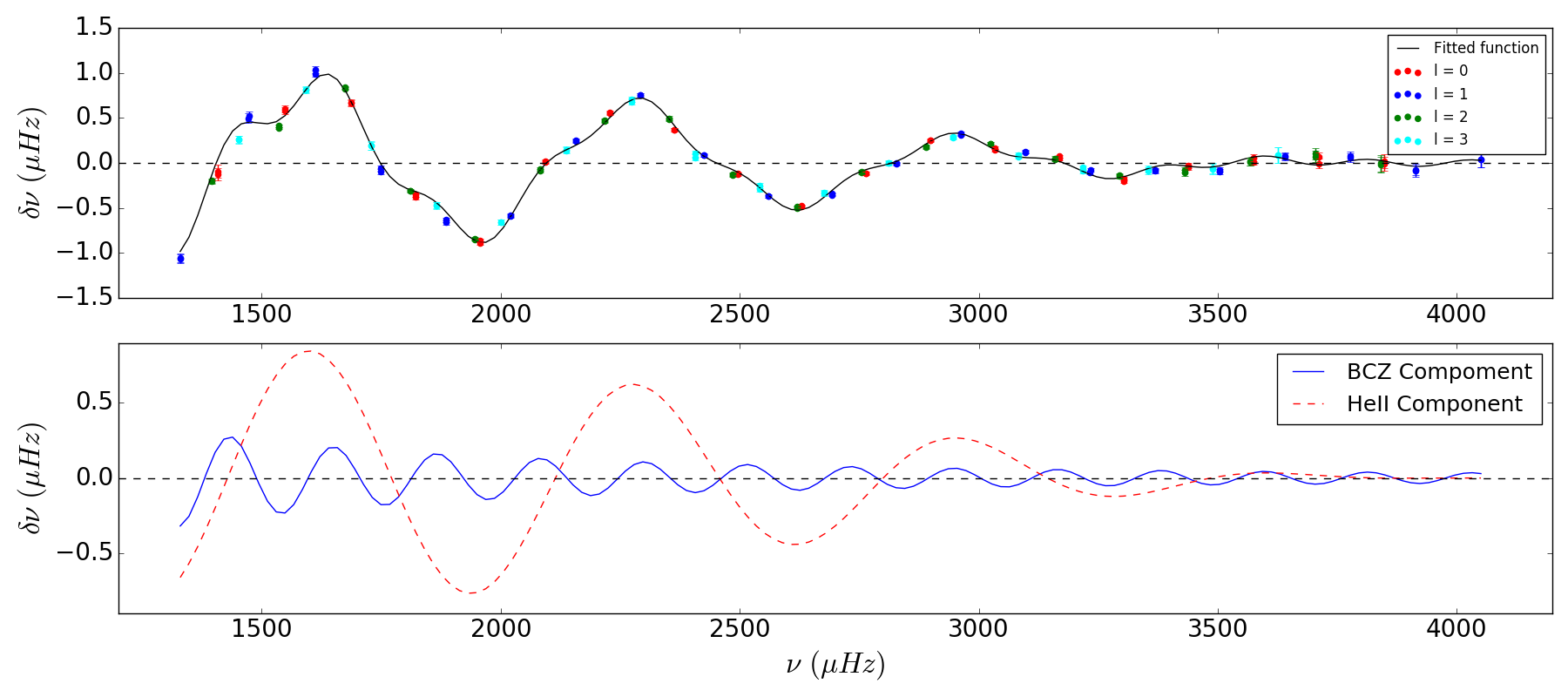}
\caption{The top panel shows the fit of the residuals from the solar frequencies to Equation \ref{eq_freq} using Method A. The color scheme detailed in the legend is used to differentiate between the degrees l of the frequencies. The bottom panel has the signal components from both the BCZ and HeII resulting from the fitting of the upper panel.}
\label{fit_sun_freq}       
\end{figure}
Figures \ref{fit_sun_freq} and \ref{fit_sun_diff} show fits to the best parameters and the oscillatory signature of both glitches for methods A and B respectively.

The results obtained are successful in showing that the code is working correctly. The small difference in values compared to the literature is to be expected considering the small changes made to the methods implemented and to the more recent data used.


\section{Automatic Pipeline} 
\label{sec:automatic}

After validating the code using the Sun, the program was used to determine the parameters of a grid of models and the results were used to build a pipeline that was capable of reducing the parameter interval automatically to improve robustness to the results without requiring user interaction.

\begin{figure}[h]
\centering
\includegraphics[width=\hsize,clip]{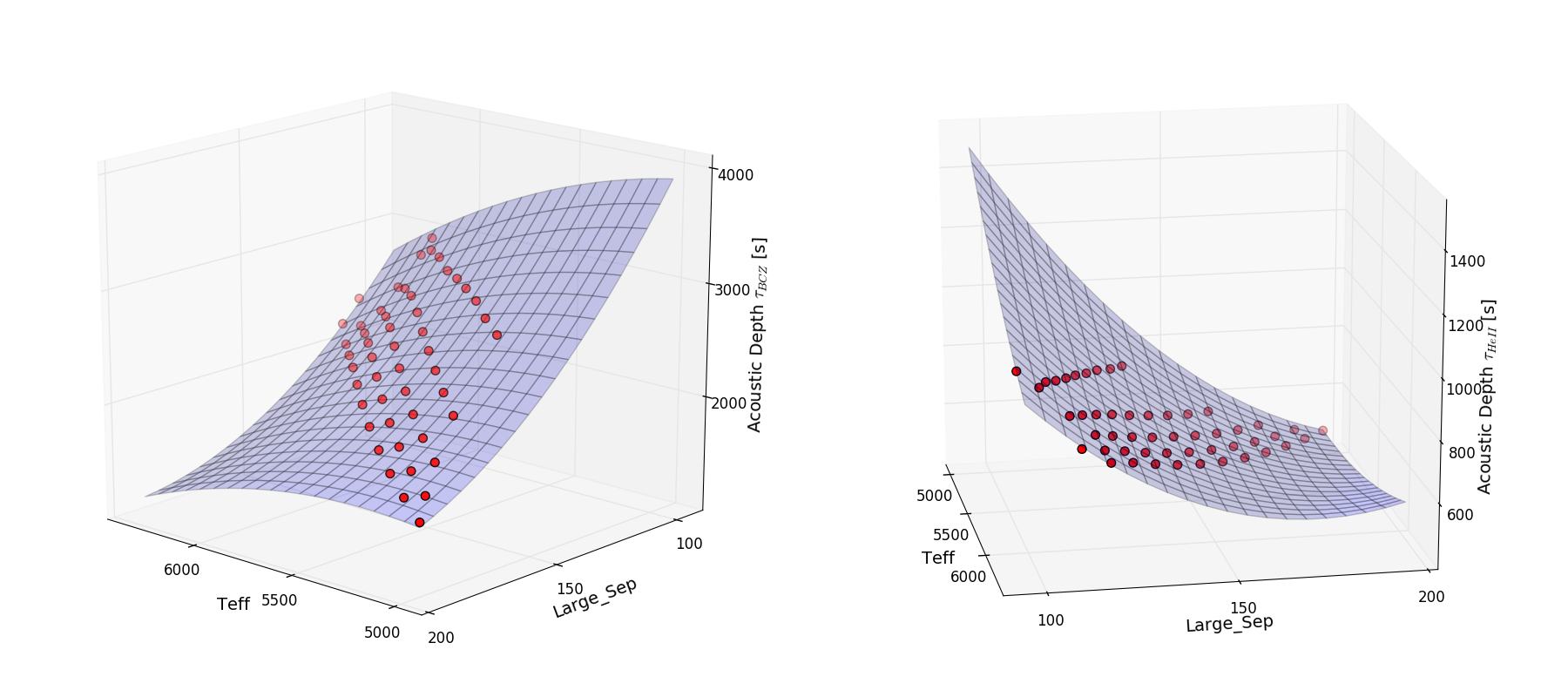}
\caption{Multivariate regression applied to the values of effective temperature, Teff, and large frequency separation, Large\_Sep, to the values of acoustic depth of the base of the convective zone (on the left) and the helium second ionization zone (on the right). The blue surface is the quadratic plane fitted.}
\label{regression}       
\end{figure}

A grid \cite{grid_cesam} of CESAM \cite{cesam} models was considered with masses between 0.8$M_\odot$ and 1.2$M_\odot$ in intervals of 0.04$M_\odot$ and each mass was considered with 5 different ages (1.5, 3, 4.5, 6 and 8 Gyr). The frequencies of the models were computed using the posc algorithm \cite{posc}. The results from this run were then used to associate the obtained values of the acoustic depths of both glitches to the values of effective temperature (Teff) and large frequency separation (Large\_sep) of the stars through a multivariate quadratic regression. The result of the regression is shown in Figure \ref{regression} . Only the results using method A were shown since the results for method B are very similar. 

The results from this regression allowed for the definition of a function that estimates the value of the acoustic depths of both glitches from the Teff and Large\_Sep of a star, improving the focus of the fitting procedure.


\section{Results} 
\label{sec:results}

To test the complete tool which is to be available to the community the program was applied to the stars from the 16 Cygni binary, with data from Verma et al. \cite{verma_2014} and to 10 solar-like stars from the sample of Mazumdar et al. \cite{mazumdar_2014} with frequencies from Appourchaux et al. \cite{star_frequencies}. The errors determined for the results were obtained by executing the program for 500 realizations of the observational data of each star, using the aforementioned Monte Carlo simulations. The results for 16 Cyg A and B are presented in Table \ref{results_cyga}. Table \ref{results_maz} shows the results for the 10 selected stars.

\begin{table}[h]
\centering
\caption{Results obtained in each method for 16 Cyg A (on the top) and 16 Cyg B (on the bottom).}
\begin{tabular}{ c c c}
\textbf{16 Cyg A} & Method A & Method B \\[0.5ex]
\hline
\rule{0pt}{1.3em} $\tau_{BCZ} (s)$ & 2904.18 $\pm$ 88.96 & 3037.64 $\pm$ 147.36 \\[1ex]
$\tau_{HeII} (s)$ & 1178.10 $\pm$ 39.11 & 981.66 $\pm$ 13.33 \\[1ex]
\end{tabular}
\label{results_cyga}

\begin{tabular}{ c c c}
\textbf{16 Cyg B} & Method A & Method B \\[0.5ex]
\hline
\rule{0pt}{1.3em} $\tau_{BCZ} (s)$  & 2538.23 $\pm$ 262.98 & 2451.79 $\pm$ 242.43 \\[1ex]
$\tau_{HeII} (s)$ & 947.26 $\pm$ 30.32 & 901.60 $\pm$ 30.61 \\[1ex]
\end{tabular}
\label{results_cygb}
\end{table}

\begin{table*}
\centering
\caption{Results obtained from both methods for the 10 selected solar-like stars.}
\begin{tabular}{ c | c | c | c | c}
& \multicolumn{2}{c}{Method A} & \multicolumn{2}{c}{Method B} \\
& $\tau_{BCZ} (s)$ & $\tau_{HeII} (s)$ &$\tau_{BCZ} (s)$ & $\tau_{HeII} (s)$ \\[0.5ex]
\hline
\rule{0pt}{1.3em}KIC008006161 & 2140$\pm$68.61 & 612.65$\pm$26.22 & 2233.96$\pm$50.12 & --- \\[1ex]
KIC008379927 & --- & 968.52$\pm$30.29 & --- & 840.52$\pm$72.51 \\[1ex]
KIC008760414 & --- & --- & --- & --- \\[1ex]
KIC006603624 & --- & 1083.66$\pm$51.83 & 3065.75$\pm$212.36 & 935.96$\pm$49.44 \\[1ex]
KIC010454113 & --- & 837.88$\pm$9.75 & 2599.24$\pm$89.32 & 751.12$\pm$19.48 \\[1ex]
KIC006106415 & --- & --- & --- & 962.53$\pm$80.31\\[1ex]
KIC010963065 & 2852.93$\pm$100.45 & 1006.18$\pm$32.95 & 2791.91$\pm$138.39 & 1002.78$\pm$72.13 \\[1ex]
KIC006116048 & --- & 1166.17$\pm$35.95 & --- & 1070.51$\pm$80.38 \\[1ex]
KIC004914923 & 3548.15$\pm$36.44 & 1114.29$\pm$25.13 & 3561.37$\pm$54.85 & ---\\[1ex]
KIC012009504 & --- & 1138.52$\pm$28.44 & --- & 1115.82$\pm$63.23 \\[1ex]
\end{tabular}
\label{results_maz}
\end{table*}


\section{Discussion and Conclusions} 
\label{sec:conclusions}

Regarding the stars from the 16 Cygni binary, both show precise results that are very similar with the ones found in other works \cite{verma_2014}. However, for the 10 selected stars the results are considerably poorer with some stars having no convergence achieved with any of the methods. Nevertheless, the results are very similar to those found in Mazumdar et al. \cite{mazumdar_2014} which might indicate that the poor results are related to the amount of available data for these 10 stars and not to the methods adopted, since these are proven to work for stars with more data. 

To conclude, from the results obtained it seems that, even with the automatic pipeline that reduces the parameter space for the fitting algorithm, the methods are highly dependent on the amount and quality of the available data. Still, the results for the Sun and the stars of the 6 Cygni binary show that is tool has potential as it achieves the same results as other methods without the need for user intervention and guidance. There are also some improvements that can be done to the program such as increasing the number of free parameters to the fitted functions to try and constrain the signal better.






\section{Acknowledgements} 
\label{sec:acknowledgements}
\textit{Part of this work had the support of the European Comission under the SPACEINN project (FP7-SPACE-2012-312844) and from FCT (UID/FIS/04434/2013).}


{\small 
\bibliography{FilipePereira}}
%
%

\end{document}